# *Optical Tunneling through Arbitrarily-Shaped Plasmonic Channels and Sharp Bends*

Andrea Alù, and Nader Engheta[*]

Department of Electrical and Systems Engineering

University of Pennsylvania Philadelphia, PA 19104, U.S.A.

E-mail: andreaal, engheta@ee.upenn.edu

**Abstract**

We propose a mechanism for optical energy squeezing and anomalous light tunneling through arbitrarily-shaped plasmonic ultranarrow channels and bends connecting two larger plasmonic metal-insulator-metal waveguides. It is shown how a proper design of sub-wavelength optical channels at cut-off, patterned by plasmonic implants and connecting larger plasmonic waveguides, may allow enhanced resonant transmission, inspired by the anomalous properties of epsilon-near-zero (ENZ) metamaterials. The resonant tunneling is shown to be only weakly dependent on the channel length and its specific geometry, such as possible presence of abruptions and bends.

PACS Numbers: 78.66.Sq, 42.82.Et, 52.40.Db, 52.40.Fd

---

[*] To whom correspondence should be addressed. E-mail: engheta@ee.upenn.edu

# 1. Introduction

Transmission enhancement of light through individual holes or arrays of nanoapertures in plasmonic screens has been recently shown theoretically and experimentally (see, e.g., [1]-[6] and references therein). Several different phenomena are possibly involved, among which surface plasmon polaritons supported by the screen [3], leaky-wave excitation [1]-[3], Fabry-Perot resonant tunneling [4]-[5] and resonant matching at the two sides of the screen [6]. Different optical applications may take advantage of such anomalous localized transmission, e.g., subwavelength imaging, nanolithography and energy harvesting.

Here we propose a fundamentally different mechanism that may induce dramatic light squeezing and transmission enhancement through arbitrarily shaped ultranarrow channels connecting larger plasmonic waveguides, with the advantages of minimized reflection, uniform phase and uniformly enhanced optical field amplitude along the channel. Such anomalous light tunneling is inspired by our recently studied resonant transmission achievable through sub-wavelength narrow channels filled by metamaterials with permittivity near zero (ENZ) at microwave frequencies [7]-[12]. Their large phase velocity, in fact, has been shown to ensure an anomalous wave tunneling and supercoupling through ultranarrow channels, with low phase delay between entrance and exit faces and uniform field across the channel length -- properties that are in principle independent of the channel specific geometry, despite possible bending, sharp abruptions or length variations. Similar features may be of great appeal for several optical applications, spanning sub-wavelength imaging, power extraction, optical communications, light squeezing and molecular fluorescence enhancement.

As it was discussed and experimentally verified at microwave frequencies in [8]-[11], even narrow hollow waveguide channels may provide analogous effective ENZ tunneling properties, exploiting the intrinsic modal dispersion in waveguides. Operated near its cut-off frequency, a microwave waveguide indeed supports a large phase velocity, in many ways analogous to a zero-permittivity (ENZ) material [13]-[16], without the need for employing specifically tailored complicated inclusions for realizing the ENZ metamaterial. Here, we apply and extend these concepts to optical plasmonic waveguides, introducing a conceptually novel resonant mechanism for light transmission and tunneling with low phase delay and uniform amplitude across arbitrarily shaped narrow plasmonic channels connecting two larger optical waveguides. Our analysis shows the inherent possibility of achieving low reflection and reasonably high transmission peaks, despite material absorption, frequency dispersion and large geometry mismatch, in sub-wavelength channels with arbitrary shapes connecting regular plasmonic metal-insulator-metal optical waveguides. We assume in the following an $e^{-i2\pi ft}$ time convention.

## 2. Geometry and Cut-Off Dispersion

Consider the geometry of Fig. 1, consisting of a parallel-plate metal-insulator-metal plasmonic waveguide, infinitely extent in the $x$ and z directions and of height $a$ supporting its dominant $TM$ mode that propagates along the $z$ axis, as indicated in the figure. In the following discussion, we consider the waveguide to be made of silver, following a Drude model $\varepsilon_{Ag} = \varepsilon_0 \left( \varepsilon_\infty - \frac{f_p^2}{f(f+i\gamma)} \right)$, with $f_p = 2175\,THz$, $\gamma = 4.35\,THz$ and $\varepsilon_\infty = 5$, which is close to experimentally measured bulk permittivity of silver in the frequency range of interest [17]. Without loss of generality, here the insulator gap is assumed to be free space with permittivity $\varepsilon_0$.

The region $|z| < l_{ch}/2$ of the waveguide has been mostly filled by silver, apart from a narrow air gap of height $a_{ch} \ll a$ in the bottom part of the waveguide. This gap is patterned by a periodic arrangement of thin silver implants of width $t$, running all over the channel. The air gap between two neighboring implants is of width $b > t$. Fig. 1 shows one unit cell of this channel geometry, implying a periodic arrangement in the $x$ direction.

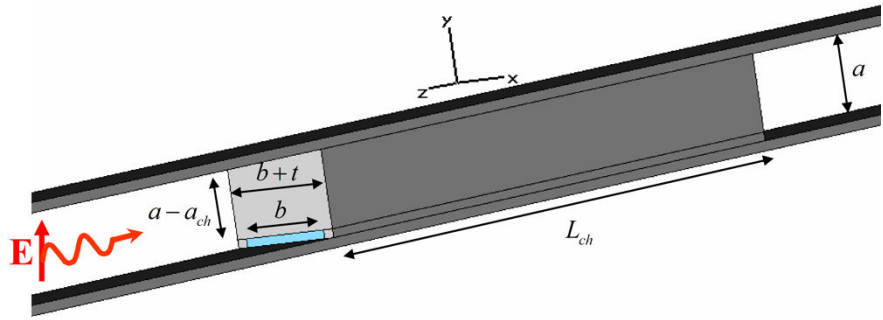

Figure 1 – (Color online) Geometry of a plasmonic ultranarrow channel connecting two larger plasmonic (metal-insulator-metal) waveguide sections (unit cell in the $x$ direction).

Consider now the transmission of the guided wave through such array of narrow channels. Since silver is a plasmonic material, with negative real part of permittivity and moderate losses, the region $|z| < l_{ch}/2$ is mostly opaque at IR and visible frequencies and therefore major part of the impinging guided optical energy carried by the guided surface plasmon wave is expected to be reflected back at the entrance of the channels. However, in [8]-[11] it was shown that for a similar waveguide geometry made of conducting materials at microwave frequencies a proper design of the sub-wavelength ultranarrow rectangular channel connecting two larger waveguide sections may lead to dramatic transmission enhancement and energy squeezing, despite the small height of the channel. This may be obtained provided that the lateral width $b$ of the rectangular channel is at the cut-off of its dominant $TE_{10}$ mode, which happens at microwave frequencies when its lateral dimension $b = \lambda_0 / 2$, where $\lambda_0$ is the wavelength in the material filling the waveguide. It was shown how this channel, due to its infinite phase velocity at the cut-off frequency, would behave as if filled by an epsilon-near-zero (ENZ) material, allowing nearly total transmission and low phase delay, weakly dependent on its geometry and total length.

Here, we may obtain an analogous optical tunneling effect when we design the width $b$ of each narrow slit in the obstructed region of the waveguide of Fig. 1 such that it operates at the cut-off of its dominant mode. The problem, however, is complicated by the fact that the ideal $TE_{10}$ mode supported by a rectangular metallic waveguide is now strongly perturbed at visible frequencies [18]-[19], since metals lose their conducting properties at these high frequencies, possessing finite negative permittivity and non-negligible losses. The modal dispersion of such plasmonic ultranarrow 4-wall channels may be evaluated by considering the coupling between the *TM* mode supported by the upper and lower plates of the channel and the *TE* mode confined between the lateral walls [18]-[19].

The actual modal distribution supported by this rectangular plasmonic waveguide is hybrid in nature and, to the best of our knowledge, no exact closed-form analytical solution is available. However, since the permittivity of silver is sufficiently negative at IR and optical frequencies, the modal distribution is not expected to be qualitatively much different from that of an ideal $TE_{10}$ mode supported by perfectly conducting walls. As we show in the following, however, its wave number $\beta$ may be drastically affected by the finite conductivity and losses of optical metals.

The parallel-plate waveguide formed by the upper and lower walls supports an even mode, of interest here, whose frequency dispersion may be evaluated by solving the dispersion equation [20]:

$$\tanh\left[\sqrt{\beta_{pp}^2 - k_0^2}\,\frac{a_{ch}}{2}\right] = -\frac{\varepsilon_0}{\varepsilon_{Ag}}\frac{\sqrt{\beta_{pp}^2 - k_{Ag}^2}}{\sqrt{\beta_{pp}^2 - k_0^2}}, \tag{1}$$

where $k_0 = 2\pi/\lambda_0$ is the wave number in the filling material (free space in our geometry) and $k_{Ag}$ is the wave number in silver (which is almost purely imaginary, since the silver, possessing permittivity with a negative real part, is opaque at the frequency of interest). As is well known, the value of $\beta_{pp}$ in a plasmonic parallel-plate waveguide with $a_{ch} \ll \lambda_0$ may become substantially larger than the ideal case of a waveguide made of ideally conducting material, for which $\beta_{pp} = k_0$.

The value of $\beta_{pp}$ from Eq. (1) is perturbed by the presence of the lateral metallic implants of thickness $t$ and periodicity $b+t$ inside the subwavelength channel of Fig. 1. Provided that $t$ is larger than the skin depth $\delta = \text{Im}\left[k_{Ag}\right]^{-1}$ of silver, the implants form a transverse array of rectangular plasmonic channels of width $b$, weakly coupled with each other, whose filling material may be effectively treated as supporting the wave number $\beta_{pp}$, consistent with effective index techniques [18]. The dispersion of the dominant quasi-$TE_{10}$ mode of such rectangular plasmonic waveguide, therefore, may be approximately calculated by considering the finite transverse impedance of its lateral walls, given by $Z_{Ag} = \sqrt{\mu_0/\varepsilon_{Ag}}$, providing the following dispersion equation:

$$\tan\left[\sqrt{\beta_{pp}^2 - \beta^2}\,\frac{b}{2}\right] = \frac{\sqrt{\beta^2 - k_{Ag}^2}}{\sqrt{\beta_{pp}^2 - \beta^2}}, \qquad (2)$$

where $\beta$ is the modal wave number, of interest here, supported by the rectangular plasmonic waveguide. The cut-off frequency of the rectangular channel, in particular, is obtained when $\text{Re}[\beta] = 0$. In the limit $\varepsilon_{Ag} \to -\infty$, the previous equations correctly converge to the correct cut-off value $b \to \lambda_0/2$ and $\beta_{pp} \to k_0$ for the perfectly conducting scenario.

Since the dominant mode supported by this rectangular plasmonic waveguide is quasi-$TE_{10}$ in the limit of $\varepsilon_{Ag} \ll -\varepsilon_0$, we may define an effective relative permittivity "seen" by the mode, generalizing the result presented in [13] for the case of ideal conductors. In this more general case, its value may be given by:

$$\frac{\varepsilon_{eff}}{\varepsilon_0} = \frac{\beta^2}{k_0^2} = \frac{\beta_{pp}^2}{k_0^2} - \frac{\pi^2}{\left(\beta_{pp} b + \frac{2k_0}{|k_{Ag}|}\right)^2}, \qquad (3)$$

which effectively takes into account the dispersion of the four plasmonic walls, and it applies in the limit of small losses. The phase velocity $2\pi f / \text{Re}[\beta]$ inside the rectangular waveguide becomes very large when $\text{Re}[\varepsilon_{eff}] \to 0$, condition for the cut-off and the effective ENZ operation of the channel.

In order to highlight the influence of the four plasmonic walls in this scenario, as compared with the perfectly conducting case, Figure 2 shows the frequency dispersion of the required width $b$ versus frequency in order to achieve cut-off at the desired frequency $f$ for three different scenarios: an ideal rectangular waveguide closed by perfectly conducting walls (blue line), a waveguide whose lateral sides are ideal conductors, but with silver on top and bottom (red) and the case of Fig. 1, where all the waveguide walls are made of silver (black). In all cases $a_{ch} = 20\,nm$ has been assumed. It may be clearly seen how the presence of plasmonic walls significantly shortens the effective wavelength in the waveguide, reducing considerably the required width to achieve the cut-off [18]. This effect is to be taken into account for a proper design of the ENZ tunneling, as we describe in the following.

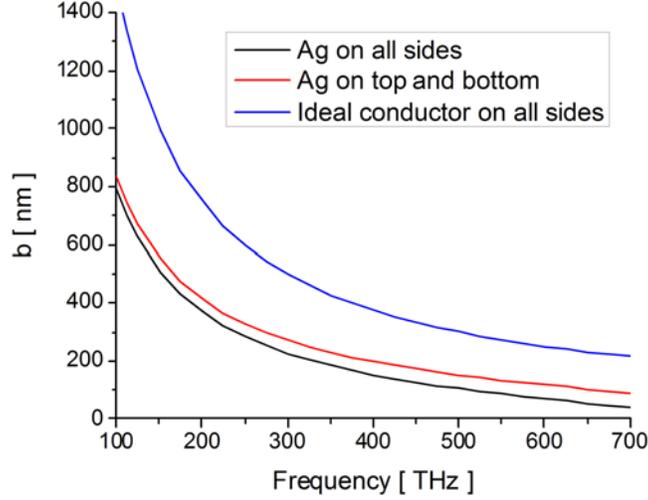

Figure 2 – (Color online) Calculated cut-off width $b$ versus frequency for the rectangular plasmonic waveguide formed by the narrow channel of Fig. 1 for $a_{ch} = 20\,nm$. The black curve refers to Eq. (2), calculated assuming that the waveguide has four walls made of realistic silver, the red line neglects the presence of silver on the lateral walls, the blue line refers to the case of ideal perfectly conducting walls for a rectangular waveguide of the same size.

In order to further clarify the modal dispersion features in such sub-wavelength rectangular plasmonic channels, Fig. 3 reports the frequency dispersion of the modal wave number $\beta$ for a rectangular channel with $b = 200\,nm$ and $a_{ch} = 20\,nm$. In the figure, the dispersion curve (black line) is compared to that of a parallel-plate plasmonic (metal-insulator-metal) waveguide with same thickness $a_{ch}$ (red line). It is evident that around $350\,THz$ the rectangular plasmonic channel has its cut off, for which $\text{Re}\left[\varepsilon_{eff}\right] = \text{Re}\left[\beta\right] \simeq 0$. The imaginary part of $\beta$, reported in Fig. 3b, suggests that in the propagation regime the absorption levels are just slightly higher than those of a parallel-plate plasmonic waveguide composed of the same material. This is a general feature of these plasmonic channels, and it is associated with the intrinsic metal absorption at optical and IR frequencies.

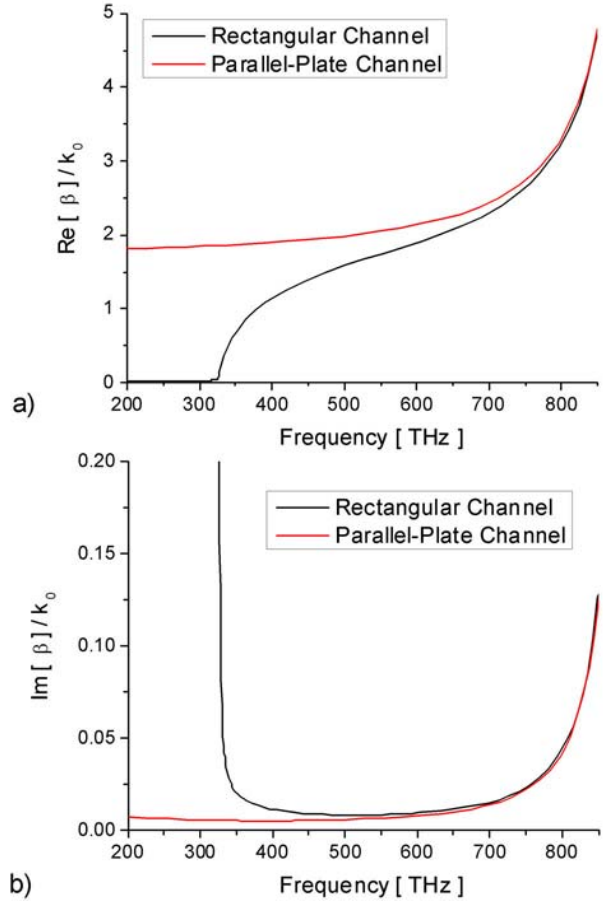

Figure 3 – (Color online) Dispersion of the guided wave number (real and imaginary parts) in a rectangular plasmonic channel with $a_{ch} = 20\,nm$ and $b = 200\,nm$ (black), compared with the dispersion in a parallel-plate metal-insulator-metal waveguide with $a_{ch} = 20\,nm$ (red).

Figure 4 reports the variation of the cut-off frequency versus $a_{ch}$ (Fig. 4a) and versus $b$ (Fig. 4b), fixing the other dimension of the channel. It is noticed how, by increasing $b$, the cut-off frequency correspondingly decreases, as expected, consistent with the behavior of a regular rectangular waveguide made of perfect conductors. However, a reduction in the vertical size $a_{ch}$ surprisingly decreases the cut-off frequency, consistent with the fact that $\beta_{pp}$ is increased at a fixed frequency by a reduction of the distance between the parallel plates in a metal-insulator-metal waveguide [20].

Figure 2-4 may be utilized for tailoring the cut-off frequency of the rectangular channels of Fig. 1, in order to tune the ENZ tunneling frequency supported by the channel, as described in the following section.

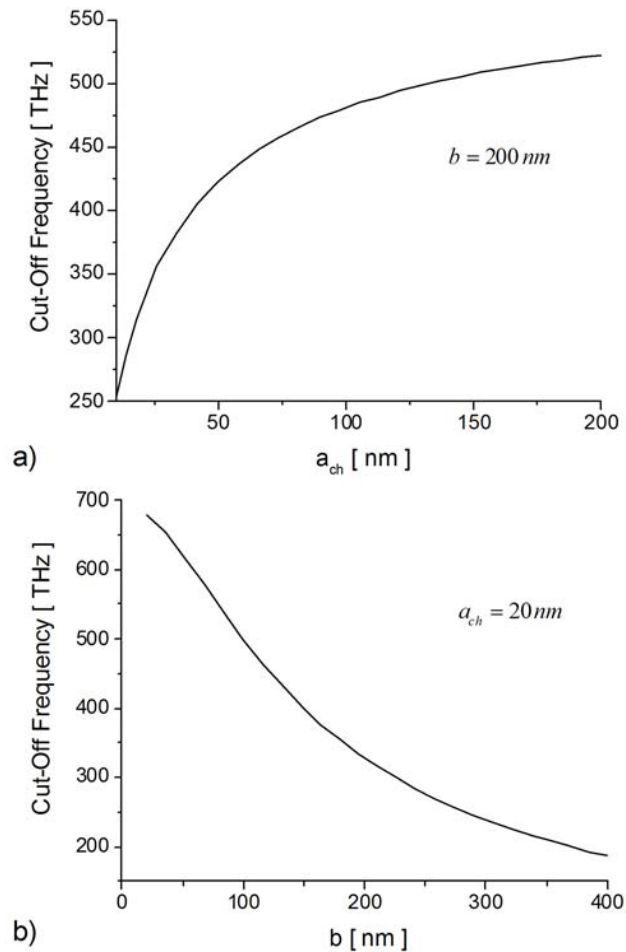

Figure 4 – Calculated dispersion of the cut-off frequency vs $a_{ch}$ for fixed $b = 200\,nm$ (a) and vs $b$ for fixed $a_{ch} = 20\,nm$ (b).

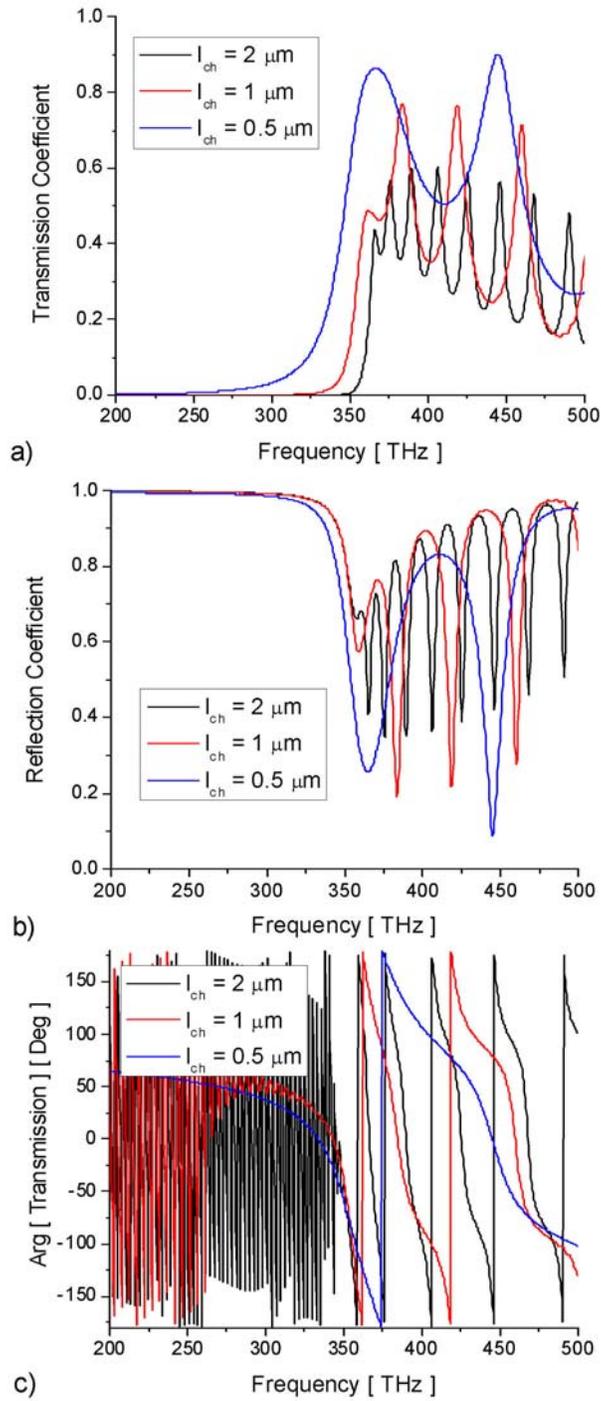

Figure 5 – (Color online) Transmission (a) and reflection (b) coefficients and phase of the transmission (c) for the channels of Fig. 1 with $a_{ch} = 20\,nm$, $a = b = 200\,nm$, $t = 50\,nm$, for three different values of $l_{ch}$.

## 3. Energy Squeezing and ENZ Tunneling

We design the narrow sub-wavelength channel of Fig. 1 to have its cut-off frequency around $f = 350\,THz$, i.e., with $a_{ch} = 20\,nm$ and $b = 200\,nm$, consistent with the geometry of Fig. 3. The implant width $t = 50\,nm$ is sufficiently larger than $\delta = 24\,nm$ to ensure that the channels are not significantly coupled with each other. The outer plasmonic waveguide has a height $a = 200\,nm = 10\,a_{ch}$.

The transmission and reflection from the channels have been evaluated in amplitude and phase, as referred to the entrance and exit planes, for the geometry of Fig. 1 using finite-integration technique commercial software [21]. The results are reported in Fig. 5. It is indeed possible to obtain strong transmission resonant peaks for specific frequency values, which are responsible for energy squeezing and wave tunneling through the sub-wavelength optical channels, despite the strong geometrical mismatch and obstruction of between the outer plasmonic waveguide and the narrow rectangular channels.

In particular, Fig. 5 reports the transmission amplitude (Fig. 3a), reflection amplitude (Fig. 3b) and transmission phase (Fig. 3c) for three different lengths of the channel, i.e., $l_{ch} = 2\,\mu m$ (black line), $l_{ch} = 1\,\mu m$ (red) and $l_{ch} = 500\,nm$ (blue). A first strong transmission peak is obtained, almost independent of $l_{ch}$, around $f_0 = 370\,THz$, which is the cut-off frequency for this geometry, close to the design value obtained from the previous approximate analysis. This is related to the ENZ operation of the rectangular channels, as described above. The subsequent higher-frequency transmission peaks are Fabry-Perot resonances, strongly dependent on the length and geometry of the channel. However, the tunneling at the ENZ frequency $f_0$ relies on a very unique [7]-[11] and distinct matching phenomenon [10], possible due to the large phase velocity along the channel, which allows uniform phase distribution and enhanced field amplitude all over the channel length, essentially independent of its geometry. The transmission levels decrease with the channel length, due to increased material absorption, consistent with Fig. 3b (in the ideal lossless limit, all the transmission peaks would converge to unity). It is noticed that, due to the large electric field enhancement, these increased material losses are inevitable if one wants to squeeze energy in such a subwavelength channel (it may be shown that the electric field is enhanced inside the channels by a factor proportional to the ratio $a/a_{ch}=10$ [10]). This explains why the transmission levels are higher for a shorter channel and why the cut-off tunneling provides somewhat a lower level of transmission when compared with some Fabry-Perot resonances: due to its uniform field amplitude, the absorption along the channel may be somewhat larger. The transmission bandwidths are influenced by the length of the channel, as seen in the figure, since a longer channel implies a higher resonance Q factor.

It is interesting to note that the reflection is correspondingly minimized at the tunneling frequencies, ensuring that the impinging energy may be effectively squeezed through the sub-wavelength channels, despite the large geometrical mismatch. At the ENZ tunneling frequency, the phase difference between entrance and exit phase is very low, as evident in Fig. 5c, essentially independent of the channel length, due to the large phase velocity ($\varepsilon_{eff} = \mathrm{Re}[\beta]=0$) near the channel cut-off. Indeed, the channels at cut-off behave as if they were filled by low-permittivity metamaterials, following Eq. (3).

Figure 6 reports the electric $E_y$ and magnetic $H_x$ field distributions, snapshot in time, for the geometry of Fig. 5 with $l_{ch} = 1\mu m$ at the cut-off frequency $f_0$. The fields have been scaled logarithmically, due to the large value of the electric field in the channel, in order to make the field distribution in the outer plasmonic waveguides visible. The distributions are shown on a plane at the center of the rectangular channel.

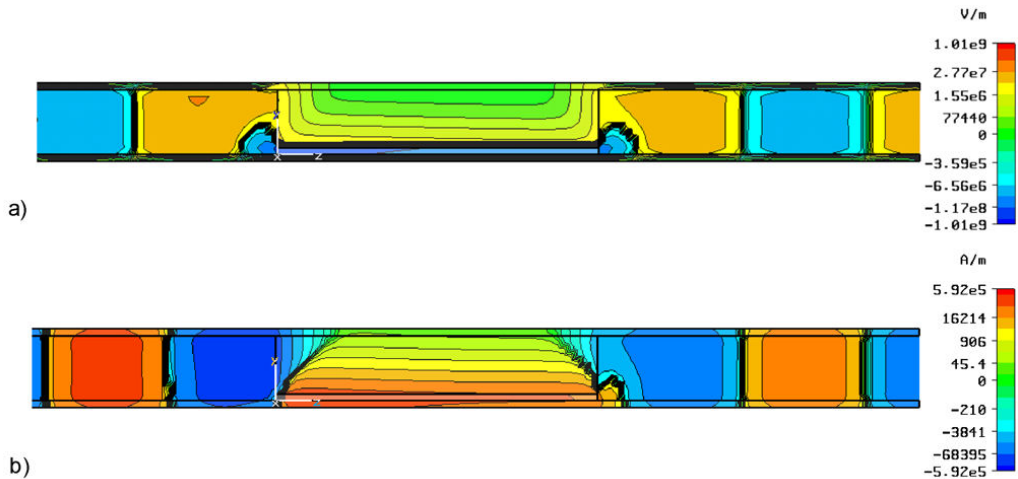

Figure 6 – Electric (a) and magnetic (b) field distributions (snapshots in time) for the geometry of Fig. 3 with $l_{ch} = 1\mu m$ at the frequency $f_0$. The distribution is on the plane at the center of the rectangular waveguide

We notice that the electric and magnetic fields are almost uniform in amplitude and phase inside the channel, despite its 1-$\mu m$ length. A good field transmission and energy squeezing is achieved through the channel, consistent with Fig. 3, and the effective wavelength in the channel is much longer than in the outer plasmonic waveguide, despite the shrinking of the channel height (which would instead tend to dramatically shorten the modal wavelength). The sharp abruption in the cross-section of the waveguide excites higher-order evanescent modes localized near the entrance and exit faces of the channel, which however do not sensibly affect the resonant tunneling and energy squeezing, again consistent with analogous findings at microwave frequencies for the case of conducting channels [7]-[11].

It is interesting to underline how the channel geometry considered here is somehow consistent with [5], in which enhanced transmission through sub-wavelength rectangular hole arrays in a plasmonic screen was demonstrated. Here, however, we are considering transmission through *long* subwavelength channels made possible by the anomalous transmission properties of effective ENZ channels, and not simply apertures in a thin screen. These results show the possibility of matching outer plasmonic waveguides through much narrower plasmonic channels of rectangular shape at the desired (and tunable) frequency of operation.

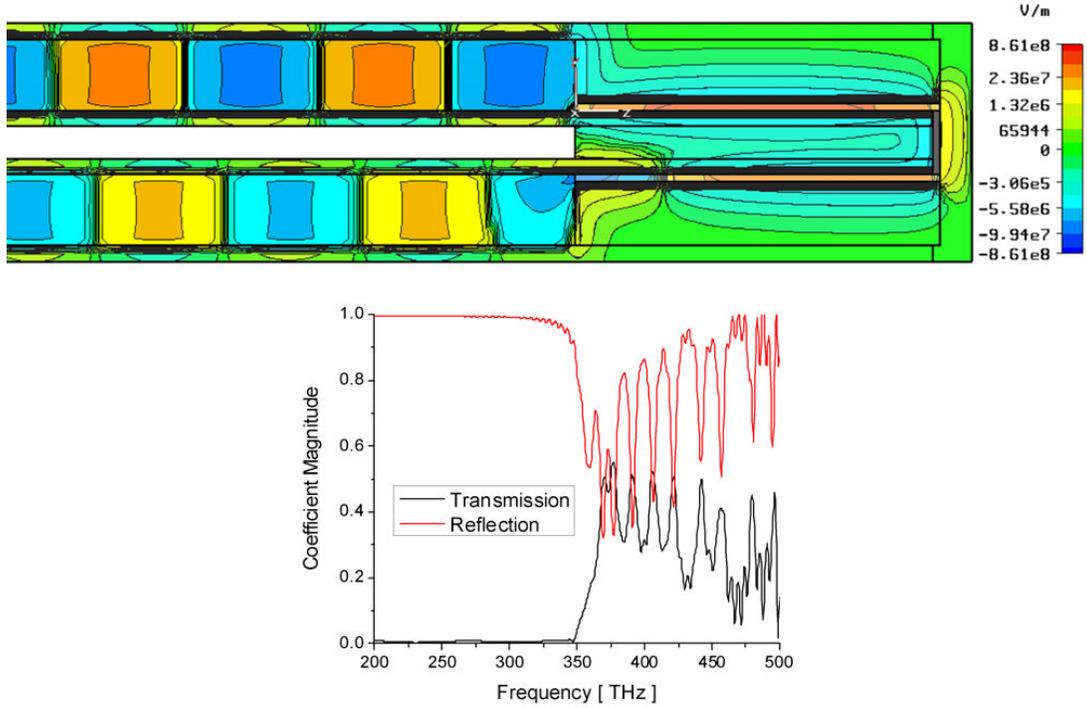

Figure 7 – Abrupt bending in the channels of Fig. 4, showing how the channel geometry does not influence the tunneling properties of the anomalous tunneling at frequency $f_0$.

A striking property of the ENZ operation of these sub-wavelength channels resides in the weak dependence of its tunneling frequency and corresponding uniform field distribution on the length of the channel, and more in general weak dependence on its geometry. This is due to the zero effective permittivity of the channel, consistent with Eq. (3), which provides "static-like" tunneling effect, independent of the specific channel geometry [7]-[11]. Consider, for instance, the geometry of Fig. 7, in which the channel of Fig. 3-4 is abruptly bent to feed a plasmonic waveguide parallel to the first one, but placed underneath. The transmission enhancement is still available at the same frequency $f_0$, with the same long-wavelength properties and analogous transmission levels. In this case, due to the total length of the channel, multiple Fabry-Perot resonances are also visible at higher frequencies, but the cut-off frequency remains around the frequency $f_0$ and the transmission is comparable with the previous examples, despite the sharp bending across the channel. In this case, the transmission for higher-order Fabry-Perot resonances is decreased by the bending, but at the effective ENZ frequency the operation is less affected by this drastic change in geometry. This is consistent with our recent experiments in metallic waveguides at microwave frequencies [11]. Also in this geometry, the tunneling frequency may be tailored at will in the design, as suggested by Eq. (3) and Fig. 2-4, by properly choosing $b$ and $a_{ch}$.

## 4. Conclusions

We have presented here a novel tunneling mechanism for energy squeezing and enhanced transmission at IR and optical frequencies. Proper tailoring of the geometry of ultranarrow plasmonic channels connecting larger plasmonic waveguides allows achieving enhanced transmission through the channel, inspired by the anomalous properties of low permittivity metamaterials. Such properties make the optical tunneling weakly dependent on the channel length and geometry, such as abruptions and bending along the channels. The results may be of interest for several optical applications, i.e., nanolithography, optical communications, optical interconnects, optical filtering, power extraction and molecular emission enhancement, exploiting the uniform phase and enhanced amplitude along the channels.

**Acknowledgement:**

This work is supported in part by the U.S. Office of Naval Research (ONR) grant number N 00014 -07-1-0622.

**References:**

[1] T. W. Ebbsen, H. J. Lezec, H. F. Ghaemi, T. Thio, P. A. Wolff, *Nature* **391**, 667-669 (1998).

[2] C. Genet, T. W. Ebbesen, *Nature* **445**, 39-46 (2007).

[3] I. I. Smolyaninov, Y. J. Hung, *Phys. Rev. B* **75**, 033411 (2007).

[4] S. M. Orbons, A. Roberts, D. N. Jamieson, M. I. Haftel, C. Schlockermann, D. Freeman, B. Luther-Davies, *Appl. Phys. Lett,* **90**, 251107 (2007).

[5] W. Jia, X. Liu, *Eur. Phys. J. B* **46**, 343-347 (2005).

[6] A. K. Sarychev, V. A. Podolskiy, A. M. Dykhne, V. M. Shalaev, *IEEE J. Quantum Electron.* **38**, 956-963 (2002).

[7] M. G. Silveirinha, N. Engheta, *Phys. Rev. Lett.* **97**, 157403 (2006).

[8] M. G. Silveirinha, N. Engheta, *Phys. Rev. B* **76**, 245109 (2007).

[9] B. Edwards, A. Alù, M. Young, M. G. Silveirinha, N. Engheta, *Phys. Rev. Lett.* **100**, 033903 (2008).

[10] A. Alù, M. G. Silveirinha, and N. Engheta, under review (2008), online at: http://arxiv.org/abs/0804.3533.

[11] B. Edwards, A. Alù, M. G. Silveirinha, N. Engheta, under review (2008), online at: http://arxiv.org/abs/0802.3540.

[12] R. Liu, Q. Cheng, T. Hand, J. J. Mock, T. J. Cui, S. A. Cummer, and D. R. Smith, *Phys. Rev. Lett.* **100**, 023903 (2008).

[13] W. Rotman, *IRE Trans. Antennas Propag.* **AP-10**, 82-94 (1962).

[14] R. Marques, J. Martel, F. Mesa, and F. Medina, *Phys. Rev. Lett.* **89**, 183901 (2002).

[15] J. D. Baena, L. Jelinek, R. Marqués, and F. Medina, *Phys. Rev. B* **72**, 075116 (2005).

[16] S. Hrabar, J. Bartolic, and Z. Sipus, *IEEE Trans. Antennas Propagat.* **53**, 110 (2005).

[17] P. B. Johnson, and R. W. Christy, *Phys. Rev. B* **6**, 4370 (1972).

[18] R. Gordon, and A. G. Brolo, *Opt. Expr.* **13**, 1933 (2005).

[19] A. Kumar, and T. Srivastava, *Opt. Lett.* **33**, 333 (2008).

[20] A. Alù, and N. Engheta, *J. Opt. Soc. Am. B* **23**, 571 (2006).

[21] CST Microwave Studio 2007, www.cst.com.